\documentstyle[aps,prl,multicol]{revtex}
\input{epsf}
\begin{document}
\draft
\preprint{\today}
\title{Intermediate liquid-solid regime near the quantum melting \\
of a two dimensional Wigner molecule}
\author{Georgios Katomeris and Jean-Louis Pichard}
\address{CEA, Service de Physique de l'Etat Condens\'e,
           Centre d'Etudes de Saclay, 91191 Gif-sur-Yvette, France}
\maketitle
\begin{abstract} 
  For intermediate values of the Coulomb energy to Fermi energy 
ratio $r_s$, the ground state of a few spinless fermions confined 
on a two dimensional torus is the quantum superposition 
of a floppy Wigner molecule with delocalized vacancies and of a Fermi 
liquid of intersticial particles. This raises the question of the existence 
of an unnoticed liquid-solid phase between the Fermi liquid and the 
Wigner crystal for fermionic systems in two dimensions. 

\end{abstract}

\begin{multicols}{2}
\narrowtext

%
%
 An outstanding question in quantum condensed matter is to know if 
the solid and the fluid can coexist in an intermediate phase 
separating the solid from the liquid. Such a phase was suggested 
\cite{andreev} if the zero point motions of certain defects 
become sufficient to form waves propagating inside the solid. The 
defects of the crystalline solid can be simple vacancy-intersticial 
pairs or more complex excitations. Their statistics coincide 
with the statistics of the particles out of which 
the solid is made. For bosons, they may form a condensate, giving rise 
to a superfluid coexisting with the solid. This supersolid phase is 
discussed in certain bosonic models \cite{batrouni}. 
For fermions, the defects may form a Fermi liquid \cite{dzyaloshinskii} 
coexisting with the solid, such that the system is neither a solid, 
nor a liquid. Two kinds of motion are possible in it; one possesses the 
properties of motion in an elastic solid, the second possesses the 
properties of motion in a liquid.

%
%
 We study such a possibility for electrons created in a two dimensional 
($2d$) heterostructure. Our motivation is threefold. First, it becomes 
possible to create very dilute $2d$ gases of electrons or holes in field 
effect devices and to vary by a gate the carrier density $n_s$ such that the 
Coulomb energy to Fermi energy ratio $r_s \propto {n_s}^{-1/2}$ can reach 
high values where the charges become strongly correlated. The studied  
devices use doped semi-conductors \cite{kravchenko} (Si-Mosfet, Ga-As 
heterostructures, Si-Ge quantum wells) or undoped organic crystals 
\cite{batlogg}. Since charge crystallization is expected  for $r_s \approx 
37$ in a clean system \cite{tanatar} and at lower $r_s$ in the 
presence of impurities \cite{chui,bwp1}, to study in those devices 
how one goes from a Fermi liquid (low $r_s$) towards a Wigner crystal 
(large $r_s$) becomes possible. Second, the observation 
\cite{kravchenko} of a metallic behavior at low temperatures in 
those systems raises the question of a possible intermediate phase, 
which should be neither a Fermi system of localized particles (Anderson 
insulator), nor a correlated and rigid solid of charges (pinned 
insulating Wigner crystal). Third, an unexplained 
intermediate regime was numerically observed \cite{bwp1,bwp2} in small 
disordered clusters. Those studies give a first threshold $r_s^F$ above 
which the Hartree-Fock (HF) approximation does not describe \cite{benenti} 
the persistent currents driven by an Aharonov-Bohm flux $\phi$. For 
$r_s < r_s^F$, the local currents are oriented at random by the impurities, 
while for $r_s > r_F$ they align \cite{bwp1,bwp2,avishai} along 
the shortest direction enclosing $\phi$. $r_s^F$ is followed by a second 
threshold $r_s^W$ where charge crystallization occurs and the persistent 
currents vanish.  This intermediate regime characterizes the ground state 
and the low energy excitations and disappears when the excitation energy 
exceeds \cite{bwp2} the Fermi energy.  Both numerics and experiments give 
an unexplained behavior restricted to temperatures smaller than the Fermi 
temperature for similar intermediate values of the ratio $r_s$. 


   Since the low energy levels do not obey \cite{bwp2} Wigner-Dyson 
statistics, the existence of non chaotic low energy collective excitations 
due to the interplay between the kinetic energy and the Coulomb 
repulsion in the clean limit can be suspected. For this reason, 
we have studied the same cluster than in Refs. \cite{bwp1,bwp2} without 
random substrate and we have observed for intermediate values of 
$r_s$ a liquid-solid regime of a type conjectured by Andreev and 
Lifshitz. The transfer of a charge from a crystal site towards some 
intersticial site creates a vacancy-intersticial pair. This costs an 
energy $\epsilon_0$, but the pair can tunnel to give rise to a 
band of delocalized excitations of width $t_0$. Near the melting point 
of the crystal, $t_0$ may exceed $\epsilon_0$ and the Wigner solid may 
co-exist with a liquid of delocalized defects. This picture is supported by 
an analysis of the ground state (GS) using a combination of Slater 
determinants (SDs) built out from plane waves and from site orbitals. The 
plane wave SDs are given by some excitations of the non interacting system 
and correspond to an excited liquid. The site SDs describe the Wigner solid 
molecule which dominates at large $r_s$. For $r_s^F \approx 9.31 < r_s 
<r_s^W \approx 27.93$ in the studied cluster, the GS is given by a quantum 
superposition of an excited liquid and of a Wigner solid. The structure of 
the low energy excitations, the errors made assuming certain truncations of 
the Hilbert space and the responses to small perturbations confirm the 
existence of two thresholds $r_s^F$ and $r_s^W$ between which unusual 
behaviors occur.  

%
%

 We consider $N$ spinless fermions free to move on a square $L \times L$ 
lattice with periodic boundary conditions (BCs). As in 
Refs. \cite{bwp1,bwp2}, we take $N=4$ and $L=6$, the size $N_H=58905$ of 
the Hilbert space being small enough to exactly diagonalize the Hamiltonian: 
\begin{eqnarray} 
\label{hamiltonian} 
H=-t\sum_{<i,j>} c^{\dagger}_i c_j +  
U \sum_{i\neq j} \frac{n_i n_j}{2 r_{ij}} 
\end{eqnarray} 
using Lanczos algorithm. $c^{\dagger}_i$ ($c_i$) creates (destroys) a 
spinless fermion in the site $i$, $t$ is the hopping term 
between nearest neighbors and $r_{ij}$ is the shortest inter-particle 
distance in a lattice with periodic BCs. $n_i=c^{\dagger}_i c_i$ and 
the interaction strength $U$ yields a Coulomb energy to Fermi energy ratio 
$r_s=U/(2t\sqrt{\pi n_s})$ for a filling factor $n_s=N/L^2=1/9$. 

%
%
 The system remains invariant under rotation of angle $\pi/2$ and under 
translations and reflections along the longitudinal $x$ and transverse 
$y$ directions. Invariance under translations implies that 
the momentum $K$ is a good quantum number which remains unchanged 
when $U$ varies. The symmetries imply that the states are fourfold 
degenerate if $K \neq 0$ and can be non degenerate if $K=0$.

%
%

 When $U=0$, the states are $N_H$ plane wave SDs  
$d^{\dagger}_{k(4)}d^{\dagger}_{k(3)}d^{\dagger}_{k(2)}d^{\dagger}_{k(1)} 
|0>$ ($d^{\dagger}_{k(p)}$ creating a particle in a state of momentum 
$k(p)= 2\pi (p_x, p_y)/L$), $|0>$ being the vacuum state  
and $p_{x,y}=1, \ldots, L$). $K_i$ and $D_i$ being the  
momentum and the degeneracy of the states of energy $E_i$, 
the GSs ($E_0(U=0)=-13 t, K_0 \neq 0, D_0=4$) are followed by two sets of 
excitations ($E_1 (U=0)=-12 t, D_1=25$) and ($E_2(U=0)=-11 t, D_2=64$). 
We denote $|K_0(\beta)>$ ($\beta=1,\ldots,4$) the SDs of energy $-13t$ 
and of momenta $K=(0,\pm \pi/3)$ and $(\pm \pi/3,0)$ and $|K_1(\beta)>$ 
the $4$ SDs of energy $-12t$ and of momentum $K_1=0$. For the non 
interacting system, the $|K_0(\beta)>$ are $4$ orthonormal GSs and the 
$K_1(\beta)$ are $4$ orthonormal first excitations, corresponding to 
a particle at an energy $-4t$ with $k(1)=(0,0)$, two particles at an energy 
$-3t$ and a fourth particle of energy $-2t$ with momenta such that 
$\sum_{j=2}^4 k(j)=0$. One has $k(2)=(0,\pm \pi/3)$, 
$k(3)=(\pm \pi/3,0)$ and $k(4)= (\mp \pi/3,\mp \pi/3)$ or $k(2)=(0,\mp 
\pi/3)$, $k(3)=(\pm \pi/3,0)$ and $k(4)=(\mp \pi/3,\pm \pi/3)$.

When $t=0$, the states are $N_H$ Slater determinants 
$c^{\dagger}_ic^{\dagger}_jc^{\dagger}_kc^{\dagger}_l |0>$ built out 
from the site orbitals. The configurations $ijkl$ correspond to the $N_H$ 
different patterns characterizing $4$ different sites of the $L \times L$ 
square lattice. If we order the configurations by the smallest distance 
$d$ between two sites, $N_d$ denoting the number of configurations 
with inter-particle spacings larger than $d$, one has $N_1=27225, 
N_{\sqrt2}=9837, N_2=2709, N_{\sqrt5}=81$ configurations out of $58905$ 
configurations. The $N_{\sqrt5}$ configurations have the smallest 
electrostatic energies, and contain 9 square configurations $|\Box_I>$ 
($I=1, \ldots, 9$) of side $a=3$ (energy $E_0(t=0) \approx 1.80 U$), 36 
parallelograms of sides ($3,\sqrt{10}$) (energy $\approx 1.85 U$) 
and 36 other parallelograms of sides ($\sqrt{10}, \sqrt{10})$ (energy $\approx 
1.97 U$). Ordering the site configurations by increasing electrostatic 
energy, those $81$ configurations are followed by $144$ configurations 
obtained by moving a single site of a square configuration by one lattice 
spacing (deformed squares of energy $\approx 2 U$).  
 
%
%

\begin{figure}
\centerline{
\epsfxsize=9cm 
\epsffile{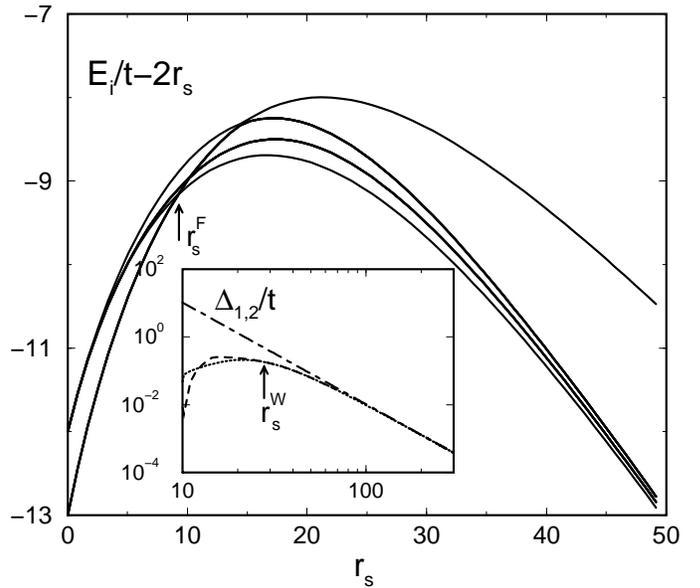}
}
\caption
{As a function of $r_s$, low energy part of the spectrum 
exhibiting a level crossing at $r_s^F$. Inset: two first level spacings 
$\Delta_1/t$ (dashed) and $\Delta_2/t$ (dotted) and the perturbative result 
$3D r_s^{-3}$ (dot-dashed).
}
\label{fig1} 
\end{figure} 

 The low energy part of the spectrum is shown in Fig. \ref{fig1} 
as a function of $r_s$. If we follow the $4$ GSs $E_0(r_s=0)$ 
($K_0\neq 0$), one can see a first level crossing with 
a non degenerate state ($K_0=0$) which becomes the GS above $r_s^F$, 
followed by two other crossings with two other sets of $4$ states 
with $K_I \neq 0$. When $r_s$ is large, $9$ states coming from $E_1(r_s=0)$ 
have a smaller energy than the $4$ states coming from $E_0(r_s=0)$. The 
degeneracies ordered by increasing energy become $(1,4,4,4)$ instead of 
$(4,25,64)$ for $r_s=0$. Since the degeneracies are $(9,36,36)$ when $t=0$, 
these $9$ states give the $9$ square molecules $|\Box_{I}>$ when 
$r_s \rightarrow \infty$. The centers of mass $R_I$ of the 
$|\Box_I>$ are located on a periodic $3 \times 3$ square lattice. 
When $r_s$ is large, one has a single massive molecule free to move on 
this restricted lattice, with a hopping term $T\propto t r_s^{-3}$ and 
quantized  $K_l(I)=2\pi p_l/3$ longitudinal and  $K_t(I)=2\pi p_t/3$ 
transverse momenta ($p_{l,t}=1,2,3$). This gives $9$ states of different 
kinetic energies given by $-2T (\cos K_l(I) +\cos K_t(I))$. The kinetic 
part of the low energy spectrum is then $-4T,-T,+2T$ with degeneracies 
$1,4,4$ respectively. This structure with two equal energy spacings 
$\Delta_1$ and $\Delta_2$ appears (inset of Fig. \ref{fig1}) when $r_s$ 
is larger than the crystallization threshold $r_s^W$. Above $r_s^W$, to 
create a defect in the rigid molecule costs a high energy available in 
the $10 ^{th}$ excitation only. 

%
%

 To describe large $r_s$, one can use degenerate perturbation theory 
and study how the degeneracy of the $9$ $|\Box_{I}>$ is removed by 
terms $\propto U/t \propto r_s^{-1}$. One obtains for the $9$ low 
energy states:
\begin{equation}
\frac {E_{I}}{t} = E_{D} -2 \frac {T}{t} (\cos K_l(I) + \cos K_t(I)).  
\label{eqDPT}
\end{equation}
 $E_{D} = A r_s + B/r_s + C/r_s^3$ and $T/t = D/r_s^3$ 
($A\approx 2.13$, $B\approx -70.81$, $C \approx -18763.48$ and 
$D \approx 3463.97$). $E_{D}$ comes from the small elastic vibrations of 
the rigid  molecule while $8T$ is the band width of its zero point 
fluctuations. 

%
%
\begin{figure}
\centerline{
\epsfxsize=9cm 
\epsffile{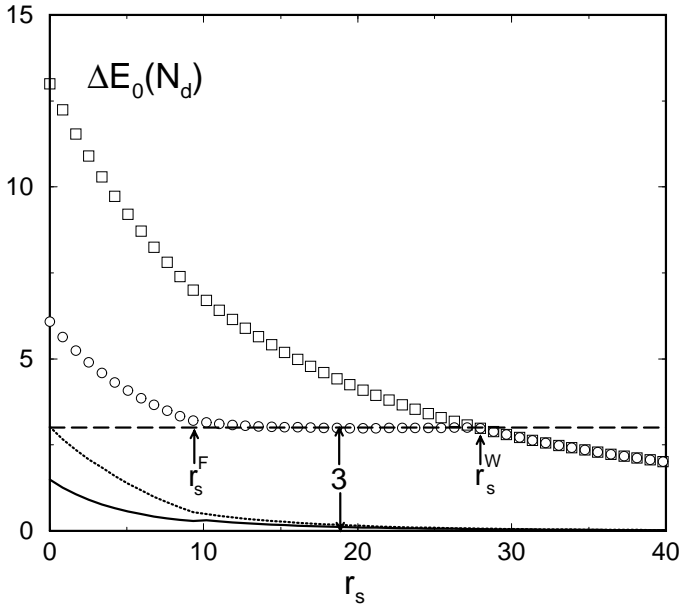}
}
\caption{ 
Errors $\Delta E_0 (N_{d}) = (E_0 (N_d)-E_0)/t$ as a 
function of $r_s$: thick line ($d=1$), dotted line ($d=\sqrt2$), 
circle ($d=2$), square ($d=\sqrt 5$). 
} 
\label{fig2} 
\end{figure} 

 The two thresholds $r_s^F$ and $r_s^W$ can be also detected if 
one calculates the GS energy $E_0 (N_{d})$ of the Hamiltonian 
truncated onto the $N_{d}$ site SDs having a minimum inter-particle 
spacing $> d$. The difference  $\Delta E_0 (N_{d}) = (E_0 
(N_d)-E_0)/t$ between the GS of the truncated Hamiltonian and 
the exact GS is given in Fig. \ref{fig2}. $\Delta E_0 (N_1)$ becomes 
negligible above $r_s^F$ where the probability to have two nearest 
neighbor particles becomes weak. $\Delta E_0(N_{\sqrt 5})$ coincides 
with $\Delta E_0 (\Box)= E_0(t=0) - E_0$. The curve $\Delta E_0 (N_2)$ 
is very remarkable. Above $r_s^W$, $\Delta E_0(N_2)$ is identical to 
$\Delta E_0 (\Box)$, while for $r_S^F < r_s < r_s^W$, $\Delta E_0(N_2) 
\approx 3$ independently of $r_s$. This suggests  that the GS for 
intermediate $r_s$ is composed of an elastic molecule, which can be 
projected onto the $N_2$ site SDs adapted to describe it, plus 
``excitations'' having a nearly interaction independent kinetic 
energy $\approx -3t$, i.e. the energy of a particle at the Fermi 
surface of the non interacting system. 
 
%
%

\begin{figure}
\centerline{
\epsfxsize=9cm 
\epsffile{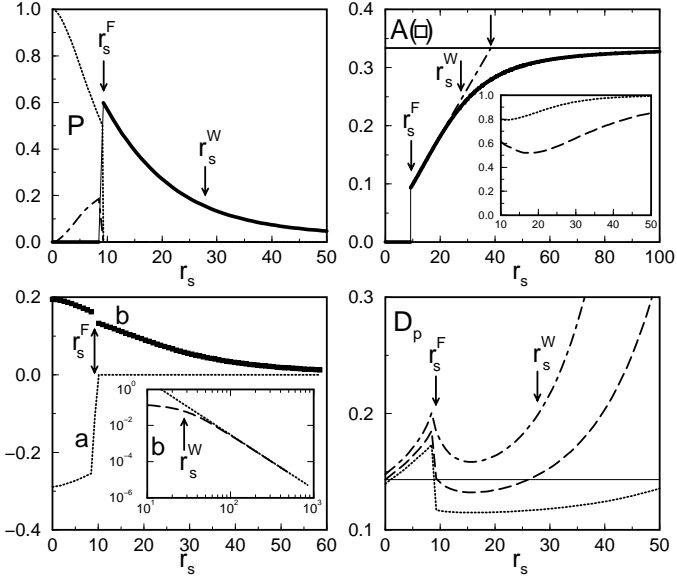}
}
\caption{ As a function of $r_s$: Upper left: GS Projections $P_O$ 
(dotted), $P_1$ (thick) and $P_2$ (dot-dashed) 
onto plane wave SDs; Upper right: GS Amplitude $A ({\Box})$ (thick) 
over a single square site SD (the dot-dashed line reaches  
$1/3$ at $r_s \approx 37$); Inset: GS Projection $p_t$ (dashed) 
and $P_T$ (dotted) over combined plane wave and site SD 
re-othonormalized bases; Lower left: $a(r_s)/t$ (dotted) and $b(r_s)/t$ 
(thick) characterizing the GS response to an infinitesimal flux; 
inset : $b(r_s)/t$ (dashed) and $8D/9 r_s^{-3}$ (dotted); Lower 
right: GS density $D_p(r_s)$ at the pinning site with 
$V_p/t=-0.01$ (dotted), $-0.05$ (dashed) and $-0.1$ (dot-dashed).  
}
\label{fig3} 
\end{figure}

 To understand further the nature of the intermediate GS, we have 
projected the GS wave functions $|\Psi_0(r_s)>$ over the two 
eigenbases valid for $U/t=0$ (Fig. \ref{fig3} upper left) 
and for $t/U=0$ (Fig. \ref{fig3} upper right) respectively. 
Below $r_s^F$, each of the 4 GSs $|\Psi_0^{\alpha}(r_s)>$ with 
$K_0\neq 0$ has still a large projection  
\begin{equation}
P_0(r_s)=\sum_{\beta=1}^4 |<\Psi_0^{\alpha}(r_s)| K_0({\beta})>|^2 
\end{equation}
over the $4$ non interacting GSs. There is no projection over the $25$ first 
excitations and a small one $P_2(r_s)$ over the $64$ second excitations of 
the non interacting system. Above $r_s^F$, the non degenerate GS with $K_0=0$ 
has a large projection 
\begin{equation}
P_1(r_s)=\sum_{\beta=1}^4 |<\Psi_0(r_s)| K_1({\beta})>|^2 
\end{equation}
which is equally distributed over the $4$ excitations $|K_1(\beta)>$ 
of momentum $K_1=0$. Its projections onto the $4$ $|K_0({\beta})>$, the 
21 other first excitations and $64$ second excitations of the non 
interacting system are zero. One concludes that a large part of the system 
remains an excited liquid above $r_s^F$, described by 4 equal projections onto 
the 4 $|K_1(\beta)>$. Those projections decrease as $r_s$ increases.  
 
 The modulus $A({\Box})=|<\Box|\Psi_0(r_s)>|$ of the amplitude of the 
projection of the GS $|\Psi_0(r_s)>$ over a single square configuration 
$|\Box>$ is given in Fig. \ref{fig3} (upper right). $A({\Box})$ does not 
depend on the chosen square and reaches $1/3$ for large $r_s$, i.e. 
the value where the total projection over the $9$ squares is equal to one. 
$A({\Box})$ linearly increases when $r_s^F < r_s < r_s^W$ 
and gives $r_s \approx 37$ (the accepted value for $2d$ Wigner 
crystallization) by extrapolation.  One concludes that a small but 
increasing part of the system begins to be a Wigner square molecule 
at $r_s^F$, described by 9 equal projections onto the $|\Box_I>$.
 The site SDs and plane wave SDs are not orthonormal. After 
re-orthonormalization, the total projection $p_{t}$ of $|\Psi_0(r_s)>$ 
over the subspace spanned by the $4$ $|K_1(\beta)>$ and $9$ $|\Box_I>$ 
and $P_t$ over the subspace spanned by the $4$ $|K_1(\beta)>$ and $225$ 
site SDs of lower electrostatic energies ($9$ squares, $36+36$ parallelograms, 
$144$ deformed squares) are given in an inset of Fig. \ref{fig3} (upper 
right). One can see that $|\Psi_0(r_s)>$ mainly remains inside this small part 
of the large Hilbert space for intermediate $r_s$. 

%
%

 We consider now the GS response to small perturbations. 
The first one consists in piercing the $2d$ torus by an infinitesimal 
positive flux $\phi$ (periodic tranverse BCs, $t \rightarrow 
t\exp(i \phi /L)$ for longitudinal hopping only, 
$\phi=\pi$ corresponding to anti-periodic longitudinal BCs).  
The coefficients $a(r_s)$ and $b(r_s)$ (Kohn curvature) of the expansion 
$E_0(r_s,\phi) \approx E_0(r_s,0) + a(r_s) \phi+ b(r_s) \phi^2 /2$ are 
given in Fig. \ref{fig3} (lower right). When $r_s=0$, $\phi$ removes the 
fourfold degeneracy of $E_0$, $a=-\sqrt{3}t/6$ and $b = 7t/36$. 
When $r_s$ is large, the substitution $K_l(I) \rightarrow K_l(I)+2\phi/3$  
in Eq. \ref{eqDPT}  gives $a=0$ and $b \approx 8Dtr_s^{-3}/9$. 
An infinitesimal positive flux $\phi$ gives rise to a persistent current 
$I_l =-\partial E_0 /\partial \phi = - a$ when $r_s < r_s^F$ while the GS 
curvature $b$ exhibits a smooth crossover between two regimes around $r_s^W$ 
(lower left inset of Fig. \ref{fig3}). 

%
%

The second perturbation consists in introducing a weak 
pinning well (negative potential $V_p$) at a single lattice 
site $p$. The GS density   
\begin{equation}
D_p(r_s)=<\Psi_0(r_s)|c^{\dagger}_p c_p |\Psi_0(r_s)>
\end{equation}
at the site $p$ is shown in Fig. \ref{fig3} (lower right). If 
$V_p=0$, $D_{p}(r_s=0)=1/9$. A weak negative value of $V_p$ 
yields a larger value for $D_p(r_s=0)$. When one turns 
on the interaction, the effect is first enhanced and enhanced Friedel 
oscillations around $p$ can be observed. At $r_s^F$, $D_p$ drops and 
the interacting GS has a weaker response to a weak pinning well than 
the non interacting GS, as shown by the curves $D_p(r_s)$  and 
by the weaker Friedel oscillations observed around $p$. When $r_s$ 
is large, $D_p$ increases again and the Wigner molecule is pinned. 
This surprizingly weak response for intermediate $r_s$ suggests that 
the system may very weakly respond to the presence of weak impurities. 

%
%
 From the study of the GS projections emerges the conclusion that a 
minimal description of the intermediate GS requires to combine the two 
limiting eigenbases. Instead of taking separately many low energy 
excitations of the $U/t=0$ limit or of the $t/U=0$ limit, one can use 
the subspace spanned by the $9$ $|\Box_{I}>$ and the $4$ $|K_1(0)>$ for 
having a large fraction of $|\Psi_0(r_s)>$ for intermediate $r_s$. 
In this sense, the GS is neither solid, nor liquid, but rather the 
quantum superposition of those two states of matter, as conjectured in  
Ref. \cite{andreev}. This suggests possible improvements of the trial 
GS to use for intermediate $r_s$ in variational quantum Monte Carlo 
approaches \cite{tanatar}. Instead of using Jastrow wave functions 
improving the plane wave SDs for the liquid or the site SDs for the solid, 
it will be interesting to study if a combination of the two describing a 
solid-liquid regime will not be better for intermediate $r_s$. This will 
confirm that an unnnoticed intermediate solid-liquid phase does exist in 
the thermodynamic limit for fermionic systems in two dimensions. This might 
help to explain the $2d$ metal observed in field effect devices. Eventually, 
our results for spinless fermions might also give a natural explanation 
for the competition between the Stoner ferromagnetism and the Wigner 
antiferromagnetism observed \cite{selva} for intermediate $r_s$ when 
the spin degrees of freedom and the disorder are included. 

%
%

We thank Boris Spivak for having drawn our attention on 
Ref. \cite{andreev} and for discussions concerning Ref. \cite{spivak}. 
This work is partially supported by a TMR network of the EU.

\end{multicols}

\end{document}